\documentclass[10pt,  twocolumn, twoside, letterpaper, conference, final]{IEEEtran}
\usepackage
[
        a4paper,% other options: a3paper, a5paper, etc
        left=1.60cm,
        left=1.60cm,
        top=1.9cm,
        bottom=3.69cm,
]
{geometry}

\usepackage{cite}
\usepackage{algorithmicx}
\usepackage[ruled]{algorithm}
\usepackage{algpseudocode}
\usepackage{mathtools}

\usepackage{amsmath}
\usepackage{amsfonts}
\usepackage{amssymb}

\usepackage{subfig}
\usepackage{graphics}
\usepackage{multirow}
\usepackage{textcomp}
\usepackage{diagbox}
\usepackage{colortbl}
\usepackage{xcolor}
\def\BibTeX{{\rm B\kern-.05em{\sc i\kern-.025em b}\kern-.08em
    T\kern-.1667em\lower.7ex\hbox{E}\kern-.125emX}}
\begin{document}

\title{Distributed DoS Attack Detection in SDN: Tradeoffs in Resource Constrained Wireless Networks}

\author{\IEEEauthorblockN{Gustavo A. Nunez Segura\IEEEauthorrefmark{1}, Arsenia Chorti\IEEEauthorrefmark{3}, and Cintia Borges Margi\IEEEauthorrefmark{1}}
\IEEEauthorblockA{\IEEEauthorrefmark{1}Escola Polit\'{e}cnica, Universidade de S\~{a}o Paulo, S\~{a}o Paulo, Brazil} 
\IEEEauthorblockA{\IEEEauthorrefmark{3}ETIS UMR8051, CY Université, ENSEA, CNRS, F-95000, Cergy, France}
Email: \{gustavoalonso.nunez, cintia\}@usp.br,
        arsenia.chorti@ensea.fr
}

\maketitle

\begin{abstract}
The Software-defined networking (SDN) paradigm centralizes control decisions to improve programmability and simplify network management. However, this centralization turns the network vulnerable to denial of service (DoS) attacks, and in the case of resource constrained networks, the vulnerabilities escalate. The main shortcoming in current security solutions is the tradeoff between detection rate and complexity. In this work, we propose a DoS attack detection algorithm for SDN resource constrained networks, based on recent results on non-parametric real-time change point detection, and lightweight enough to run on individual resource constrained devices. Our experiment results show detection rates and attacker identification probabilities equal or over $0.93$.
\end{abstract}

\begin{IEEEkeywords}
Software-defined networking, intrusion detection, wireless sensor networks
\end{IEEEkeywords}

%%%%%%%%%%%%%%%%%%%%%%%%%%%%%%%%%%%%%%%%%%%%%%%%%%%%%%%%%%%%%%%%%%%%%%%%%%%%%%%%
\section{Introduction} \label{sec:intro}

Software-defined networking (SDN) is a paradigm that centralizes network control decisions and enables the network to be intelligently and centrally programmed. These characteristics simplify network management and provide tools for infrastructure sharing \cite{McKeown2008}.

SDN centralization provides advantages and disadvantages in terms of network security: on one side, the controller's global view has been used to develop new security strategies \cite{7226783}, on the other side, the controller is a single point of failure, which turns SDN-based networks prone to DoS attacks \cite{SINGH2020509} \cite{7593247}. In the case of resource constrained networks, as wireless sensor networks for Internet of things, SDN vulnerabilities are critical since there are less resources to detect and mitigate attacks. Consequently, current standard SDN security solutions adaptation is not trivial.  

Since SDN centralizes the control logic of the network, most of works in the literature propose centralized security solutions. This has benefits, such as a global view of the network and high processing power, but it also requires a constant communication between the network devices and the controller. This means more energy and communication resources consumption. To address this issue, we propose a lightweight DoS attack detection algorithm using change point (CP) analysis to detect anomalies in the network behavior. We execute our proposal using a distributed approach, running the detection algorithm on individual resource constrained nodes, avoiding the packets overhead caused by the centralization.

%In previous works we proposed a centralized DoS attack detection for SDN resource constrained networks based on change point (CP) detection. The results showed high detection rate and type of attack identification probability exceeding $0.89$, however, it requires all nodes report performance metrics to a centralize entity, i.e. management sink or SDN controller. 

We simulate grid topologies of 100 nodes, where 10\% of nodes are attackers. Our main results show that individual nodes can detect a DoS attack and identify the attacker itself with a probability equal or over $0.93$, when being close to the attacker. In addition, we investigate trade-offs between a fully decentralized and a hybrid approach.

%%%%%%%%%%%%%%%%%%%%%%%%%%%%%%%%%%%%%%%%%%%%%%%%%%%%%%%%%%%%%%%%

\section{Related work}

In this section, we analyze recent works that propose DoS attacks detection solutions for SDN-based networks. We compare our proposal and the state mainly based on detection performance and resources limitations. 

Machine learning is a popular approach used for security in SDN since the controller has access to traffic information that could be used to train the algorithms. Bhunia and Gurusamy \cite{8215418}, Ravi and Shalinie \cite{8993716}, and Jia \textit{et al.} \cite{9090824} proposals have in common that all of them obtained high detection rate results, i.e., higher than $90\%$, using machine learning techniques. On the other hand, none of these three proposals considered resource constraints. The main reason is because these are OpenFlow-based or require high traffic of packets to monitor the network. 

Some proposals focused on resource limitations. Yin \textit{et al.} \cite{8352645}, Miranda \textit{et al.} \cite{8998393}, and Wang \textit{et al.} \cite{WANG2018119} proposed more lightweight security solutions, but at the cost of detection rates below $90\%$. However, these works proposed multiple types of attack detection and attacker identification algorithms.

Commonly, security in SDN is centralized, but nevertheless, there are distributed-based proposals in the literature. One reason to use distributed approaches is to avoid control overhead that could saturate vital control links. To address this shortcoming, Fawcett \textit{et al.} \cite{8468216} proposed Tennison, a framework for scalable network security based on multi-level flow monitoring. Distributed approaches have been used also to detect anomalies in local sub-networks \cite{9216536}

The main shortcoming in the state of the art is the tradeoff between detection rate and solution complexity. The proposals that attained high detection rate were not suited for resource constrained networks, and proposals that considered resource limitations did not attain high detection rates. In this work we propose a DoS attack detection algorithm for SDN resource constrained networks, lightweight enough to run on individual resource constrained devices. Results show a detection probability comparable to centralized proposals, but reducing packets traffic, a key shortcoming in centralized solutions.

%%%%%%%%%%%%%%%%%%%%%%%%%%%%%%%%%%%%%%%%%%%%%%%%%%%%%%%%%%%%%%%

\section{SDN Security Vulnerabilities} \label{sec:vul}

As explained in Section \ref{sec:intro}, network control centralization and planes separation are fundamental enablers of SDN programmability. On the other hand, these traits turn the network vulnerable to denial of service (DoS) attacks,

In SDNs, the attackers can reach the control plane directly through the controller(s) or through network devices. Control packets flooding attacks are common since these packets have to reach the controller to be processed, which can lead to processing and communication resources exhaustion. The attackers are also able to mislead other network devices and induce them to flood the network. 

The SDN controller needs topology information to operate. To this end, the network devices send neighborhood information to the controller for configuration and control decisions. In the case of wireless SDN networks, attackers may hear this information and use it to mislead the controller to take wrong routing decisions.  

In the case of SDN resource constrained networks, these attacks target specific characteristics. Attackers can launch control plane attacks to saturate flow tables and buffers of devices with low storage capacity. A saturated node may not have space to forward new packets or receive new routing rules. This will trigger a series of packets retransmissions, which means more energy, processing and communication resources consumption. Since the network operation depends on the controller, if this do not take actions, the network devices can exhaust all their resources.     

In a previous work \cite{OJIOT2019gnunez}, we analyzed the impact of DoS attacks in SDWSNs. One of these attacks was the false data flow forwarding (FDFF) attack.%, shown in Fig. \ref{fig:fdff_a}. 
The FDFF attack targets the controller via network's devices. First, the attacker sends data packets with unknown flow identifiers to its neighbors. The neighbors receive the packet and check the flow table to determine the action required, without success, thus they ask a rule to the controller by sending a flow rule request packet. The controller receives this packet, calculates the rule and replies sending a flow setup packet. This attack increases the control overhead and the processing overhead on the nodes in the neighborhood and on the controller. Also, after several repetitions, this attack can saturate the neighbors’ flow tables. 

% \begin{figure}
%     \centering
%     \includegraphics[width=0.35\textwidth]{Figures/FDFF.eps}
%     \caption{False data flow forwarding attack (FDFF)% message exchange
%     %: the attackers inside the network send data packets to their neighbors using random or unknown identifiers. The sensor nodes request a rule to the controller to treat this packet, the controller calculates the rule and send it to the sensor node
%     }
%     \label{fig:fdff_a}
% \end{figure}

\section{Distributed DoS Attack Detection}\label{sec:cp}

From \cite{icc-2020} and \cite{latincom-2020} we know that our CP detection algorithm, based on \cite{skaperas:hal-01997965} \cite{8835019}, is able to detect FDFF attacks with a probability over $0.96$, and identify the type of attack with a probability exceeding $0.89$. In this work, we go further and execute the CP detector in a distributed approach, this means, running on individual nodes. Our objectives are: first, to evaluate the performance of the CP detector in resource constrained devices, and second, study the tradeoff when running the detectors on every node in the network and running it in clusters.

% In subsection \ref{sec:cp_math} we explain the mathematics of the CP detector and in subsection \ref{sec:implementation} we explain the distributed DoS detector implementation.

\subsection{Change point detection} \label{sec:cp_math}

The problem formulation exploits recent results \cite{skaperas:hal-01997965,8835019} on non-parametric real-time CP detection. We adapted the hybrid offline-online proposal to an entirely online detector \cite{segura2021centralized}.

To outline the online CP algorithm, let $\{X_{n}:n \in \mathbb{N}\}$ be the time series of the metric monitored. Using Wold's theorem we can assume that, for $X_1,...,X_N$, each sample is expressed as $X_n=\mu_n+Y_n$, where $\{\mu_n, n\in \mathbb{N}\}$ is the mean of the time series and $\{Y_n:n \in \mathbb{N}\}$ is a random zero mean term, so that we can rewrite $X_{n}$ as:
\begin{equation}\label{eq:11}
X_n= \begin{cases} 
         \mu+Y_n, \hspace{11mm} n=1,\ldots,m+k^*-1 \\
\mu+Y_n+I, \hspace{5mm} n=m+k^*,\ldots 
  \end{cases}
\end{equation}
where $\mu$, $I\in\mathbb{R}^{r}$, represent the mean parameters before and after the unknown time of possible change $k^*\in\mathbb{N^*}$ respectively. The term $m$ denotes the length of an initial period assuming no change on the mean value, i.e, $\mu_1=\dots=\mu_m$. During this period, our detector ``learns'' in real-time the statistics of the observed time series, and, the mean value in particular. Finally, the statistical hypothesis test is articulated as: $H_0: I=0$, $H_1: I\neq0$.

% \begin{equation}\label{eq:13}
% \begin{aligned}
% H_0&: I=0\\
% H_1&: I\neq0.
% \end{aligned}
% \end{equation}

The online analysis is a stopping time stochastic process defined as: 
\begin{equation}\label{eq:14}
\tau{\left(m\right)}= \begin{cases} 
         \min\lbrace{l\in{\mathbb{N}}: {TS_{on}(m,l)}{\geqslant{F(m,l)}}}\rbrace,\\
\infty, \text{ if } {TS_{on}(m,l)}{<{F(m,l)}} \text{ }\forall{l\in{\mathbb{N}}},
   \end{cases}
\end{equation}
where $TS_{on}(m,l)$ is the detector, calculated online for every $l$, and $F(m,l)$ is the given threshold; with properties $\lim_{m\to\infty} Pr{\left\lbrace\tau(m)<\infty|H_0\right\rbrace=\alpha}$, 
ensuring that the probability of false alarm is asymptotically bounded by $\alpha\in\left(0,1\right)$, and, $\lim_{m\to\infty} Pr{\left\lbrace{\tau(m)<\infty|H_1}\right\rbrace=1}$, ensuring that under $H_1$ the asymptotic power is unity. Under these conditions, $F(m,l)={cv_{on,\alpha}}{g}{\left({m,l}\right)}$,
% \begin{equation}\label{eq:15}
% F(m,l)={cv_{on,\alpha}}{g}{\left({m,l}\right)},
% \end{equation}
where the critical value $cv_{on,\alpha}$ is determined from the asymptotic distribution of the detector under $H_0$ and the asymptotic behavior achieved by letting $m\rightarrow\infty$. The weight function is defined as,
\begin{equation}\label{eq:16}
g(m,l)=\sqrt{m}\left(1+\frac{l}{m}\right)\left(\frac{l}{l+m}\right)^\gamma
\end{equation}
where the sensitivity parameter $\gamma\in\left[0,1/2\right)$. 

The online algorithm uses the standard CUSUM detector \cite{FREMDT201474}, with test statistic denoted by  $TS^{ct}_{on}$. Its corresponding critical value is denoted by $cv^{ct}_{on,\alpha}$ and the stopping rule by $\tau_{ct}(m)$. The sequential CUSUM detector is denoted by $ E(m,l)=\left(\overline{X}_{m+1,m+l}-\overline{X}_{1,m}\right)$.
    % \begin{equation}{\label{onlC}}
    % E(m,l)=\left(\overline{X}_{m+1,m+l}-\overline{X}_{1,m}\right)
    % \end{equation}

The standard CUSUM test is expressed as:
\begin{equation}\label{eq:17}
TS^{ct}_{on}(m,l)={l\widehat{\Omega}^{-\frac{1}{2}}_{m}}{E(m,l)},
\end{equation}
where $\widehat{\Omega}_{m}$ is the estimated long-run covariance, defined as in (4), that captures the dependence between observations. Then, the stopping rule $\tau_{ct}(m)$, is defined as: 
\begin{equation}\label{s.t1}
\tau_{ct}(m)=\min\lbrace{l\in\mathbb{N}:\Vert{TS^{ct}_{on}(m,l)}\Vert_{1}\geq{cv^{ct}_{on,\alpha}}g(m,l)\rbrace},
\end{equation}
where the $\ell_{1}$ norm is involved to modify $TS^{ct}_{on}$ so that it can be compared to a one dimensional threshold function.   
The critical value, $cv^{ct}_{on,\alpha}$, is derived from the asymptotic behavior of the stopping rule under $H_{0}$:

\begin{align} \label{eq:18}
 & \lim_{m\to \infty} Pr {\lbrace \tau(m)<\infty\rbrace} \\
 & = \lim_{m \to\infty} Pr {\left\lbrace{\sup_{1\leqslant{l}\leqslant{\infty}}}{\frac{\Vert{TS^{ct}_{on}(m,l)}\Vert_{1}}{g(m,l)}}{>cv^{ct}_{on,\alpha}}\right\rbrace} = \alpha
%  & =Pr\left\lbrace\sup_{t\in\left[0,1\right]}{\frac{\Vert{W(t)}\Vert_{1}}{t^\gamma}}>cv^{ct}_{on,\alpha}\right\rbrace=\alpha
\end{align}

% Summarizing, the overall algorithm has 3 main steps:
% \begin{itemize}
%     \item Step 1: define the values of the quantities $m$,  $\gamma$, the confidence level $\alpha$, and set $l$.
%      \item Step 2: after collecting $m$ samples of the metric, $\Gamma(m,l)$ (\ref{eq:17}) and the weight function in (\ref{eq:16}) are calculated for every $l$ on the monitoring period to then apply (\ref{eq:18}).
%      \item If a CP is detected, the online process stops. Conversely, if the period $l$ ends, a new monitoring period is defined.
% \end{itemize}

\subsection{DoS attack detection: implementation} \label{sec:implementation}

In our distributed proposal, every node is able to monitor a time series using its own metrics and execute the CP detector algorithm, and also, each node is able to send metrics samples to a cluster head (CH). In the second case, the CH is in charge of constructing the time series of the cluster and execute the detection algorithm. Such approaches, in which hardware behavior is monitored to identify anomalies, hints to further integration with other approaches for the domain of physical layer security \cite{PHYErsi} and introducing security controls at all layers.   

A security application was programmed in every node. This application manages the sampling and the detection algorithm execution. The algorithm initiates constructing a time series of $200$ samples ($m=200$), from where it extracts the statistical information that will use during the online CP detection. When the online phase starts, the algorithm continues storing samples. In the case no CP is detected during the first $50$ samples ($l=50$), these samples are added to the $200$ samples taken before to extract new statistical information. This, process is iterated every $50$ samples, but, if a CP is detected, the application rises an alarm and informs the controller about the situation. 

Since the FDFF attack increases the message exchange activity, we decided to monitor the transmitting time on every node, i.e, the number of ticks the radio module remained turned on transmitting packets. From previous results \cite{latincom-2020}, we learned that $\gamma = 0$ and $m = 200$ maximize the detection rate. In this work we used these values as well.

The security module was implemented in C language using Contiki-3.0 \cite{Dunkels}, an operating system for WSN and IoT, and IT-SDN \cite{8805072}, an SDWSN framework developed by our reserch group. For transmitting time sampling we used Energest \cite{energest}, a power management module for Contiki. For control packets sampling we implemented a counter for this specific type of packets. 

\section{Results and Analysis}

Our analysis follows two approaches: detection performance and implementation overhead. For detection performance we analyzed the attack detection probability and attacker identification probability. For implementation, we analyzed the packets overhead and memory usage. Both scenarios were simulated on Cooja \cite{Osterlind2006}, simulating grid topologies of 100 nodes, where 10\% of nodes are attackers and emulating sky motes. 

\subsection{Detection performance}

\begin{figure}[t]
    \centering
     \subfloat[FDFF detection: individual nodes\label{fig:FDFF_individual}]{%
      \includegraphics[width=0.43\textwidth]{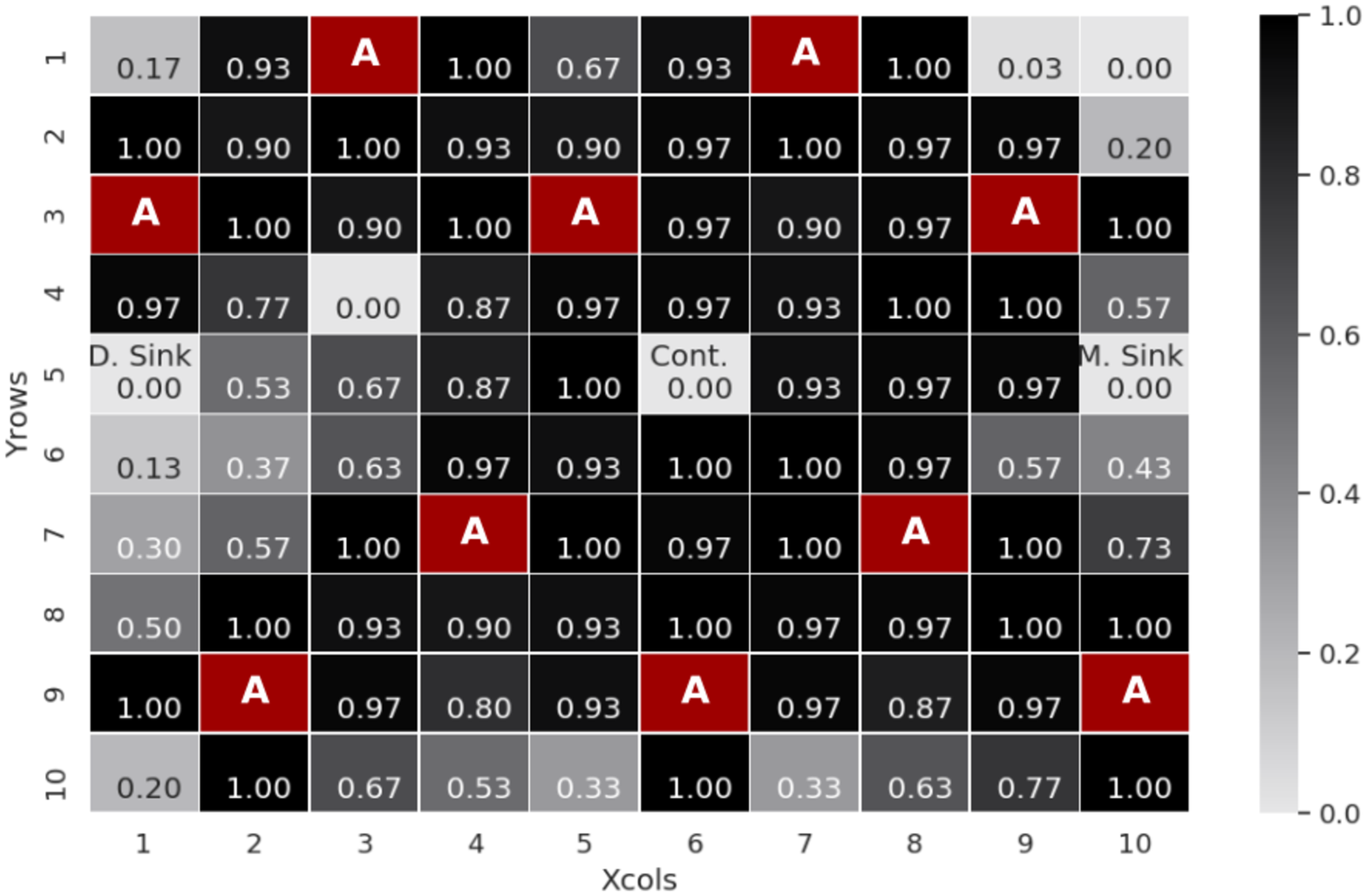}}
\hspace{\fill}
  \subfloat[FDFF detection: clusters\label{fig:FDFF_clusters}]{
  \includegraphics[width=0.43\textwidth]{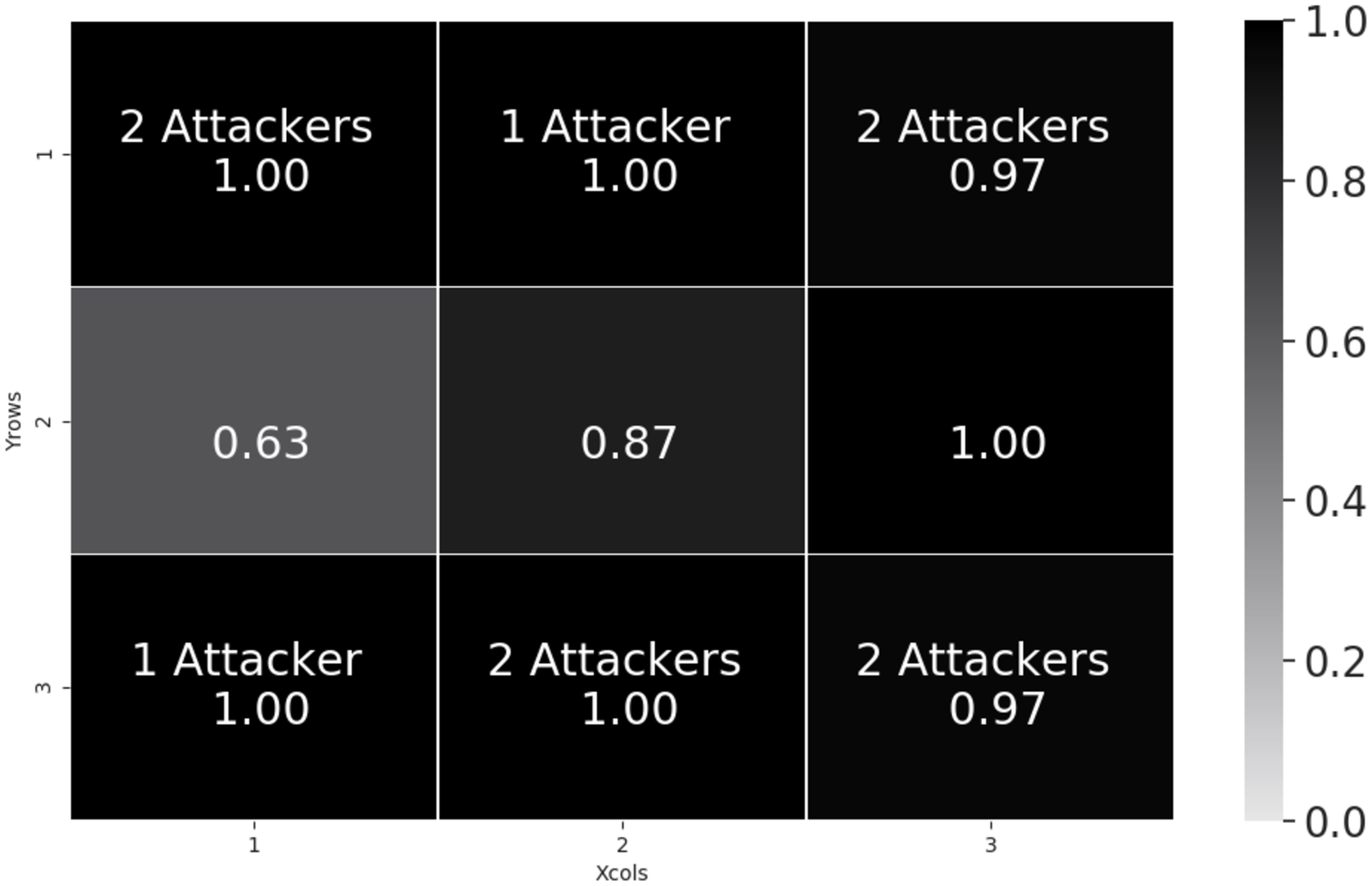}}
    \caption{Distributed FDFF attack detection%: the heatmap in Fig. a) shows the detection probability of each individual node, while the heatmap in Fig. b) shows the detection probability of each cluster
    }
    \label{fig:FDFF_prob}
\end{figure}

Fig. \ref{fig:FDFF_prob} shows the detection probability heatmap for both the distributed and hybrid approaches. For the case where each individual node is running the detector (Fig. \ref{fig:FDFF_individual}) the attackers' position is represented with an ``A''. For the case where the detector is running on clusters (Fig. \ref{fig:FDFF_clusters}), the heatmap shows also the number of attackers in each cluster.

In Fig. \ref{fig:FDFF_individual} we observed that the detection probability is higher around the attackers and around the controller, with values between $0.93$ and $1.00$. This is the behavior we expected since the FDFF attack targets the control plane through the attackers' neighbors, thus the attack has a direct impact on the transmitting time mean value of these nodes. The CPs detected on the nodes at two or more hops from the attackers reflect the impact of this attack in the whole network. The increase in the transmitting time of these nodes is caused mainly by the increase in the control packets forwarding. 

The detection probability results for the clustering case in Fig. \ref{fig:FDFF_clusters} showed that in the clusters where there are one or two attackers, the detection probability was equal or above $0.97$. In the clusters without attackers, the detection probability was between $0.63$ and $1.00$. Similar to the case running the detector on individual nodes, high detection probability results in clusters without attackers mean that this attack impacts on the whole network.

The detection probability results provided two important insights: (i) we were able to detect an FDFF attack when monitoring the transmitting time in either individual nodes or clusters, and (ii) the detection probability was lower on the nodes at two or more hops from the attackers. Based on (ii), we implemented an attacker identification algorithm for the case running on individual nodes. First, every time a node receives a data packet with an unknown flow, it saves the identification address of the sender in a vector for suspects. Then, if a node detects a CP on the transmitting time, it sends an alarm to the centralized security module and informs the address of a suspect. To determine the suspect, the node checks the last ten addresses saved in the vector and choose the one with the highest frequency. We chose ten samples because, according to \cite{latincom-2020}, the slower detection when $\gamma=0$ takes around ten samples. When the security module receives the alarm, sets a counter for every suspect. If the counter is equal to the number of neighbors of the corresponding suspect, the suspect is declared as attacker. 

\begin{figure}
    \centering
    \includegraphics[width=0.45\textwidth]{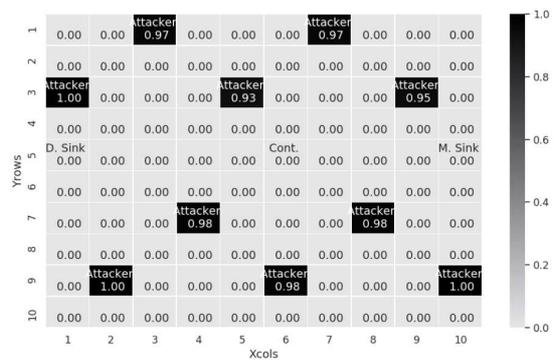}
    \caption{Attacker identification probability: }
    \label{fig:attacK_id_prob}
\end{figure}

The heatmap in Fig. \ref{fig:attacK_id_prob} showed that using this algorithm we were able to correctly identified all the attackers with a probability equal or above $0.93$. In addition, the false positives were equal to zero. On the other hand, this was possible only running the detector on every individual node. In the clustering approach, groups without attackers inside also obtained high detection rates, which excluded the possibility of tracking the attacker based only on the alarm received from the cluster.

\subsection{Implementation comparison}

In the clustering approach, the cluster head is in charge of constructing the time series for the whole cluster and execute the CP detector. To accomplish this, all nodes have to monitor their transmitting time and send a sample to the cluster head, periodically. The cluster head sums up all the samples received per period and the result is a new sample of the cluster's time series. Conversely, when the CP detector is running on individual nodes, each one constructs its own transmitting time time series and executes the CP detector.

The clustering approach has the benefit that it reduces the memory and the processing overhead of the network, since the detector is running on the cluster heads only. On the other hand, it increases the packets traffic, since all nodes now have to periodically send a transmitting time sample to the cluster head. In terms of memory, the time series construction and the CP detector implementation, for one metric, represents 5956 B. In our specific case using sky motes, this value represents 12.40\% of the total memory.

%41921	    204	   8868	  50993
%47877	    216	   9448	  57541

One security risk when using clustering approaches, is the possibility of the cluster head to be an attacker. One way to solve this is using secure cluster head selection algorithms \cite{iet-wss.2018.5008}, but this requires more memory, processing, and communication resources, which are already scarce in our case. One benefit of running the detector on every node is that we avoid this risk since the attack detection does not depend on one or few nodes only.

% \begin{figure}
%     \centering
%     \includegraphics[width=0.49\textwidth]{Figures/1-s_100.eps}
%     \caption{Detection speed:  }
%     \label{fig:1-s_groups}
% \end{figure}

\section{Conclusion}

% Software-defined networking is a paradigm devised to simplify network management, configuration and automation. Nevertheless, SDN intrinsic characteristics turn the networks prone to DoS attacks. This security problem is more intense in resource constrained networks since there are less resources to detect and mitigate security attacks. 

In this work we propose a distributed DoS attack detection for software-defined resource constrained wireless networks, based on CP detection. Our proposal is lightweight enough to run on very limited devices and detect FDFF attacks with a probability between $0.93$ and $1.00$, comparable to detection results in centralized proposals. 

This proposal was evaluated running the detector on every node and running the detector in clusters. Results showed that both approaches obtained high detection rate. Additionally, we were able to identify the attackers without increasing the packets traffic when running the detector on every node. Comparing the implementations of both approaches, the clustering approach requires less memory on every node, while running the detector on every node reduced the packets traffic. 

As future work, we envisage to develop a full implementation of distributed and centralized approaches and compare their performance based on security and network performance metrics.

\section*{Acknowledgments}\vspace{-0.1 cm}
This study was financed in part by the Coordena\c{c}\~{a}o de Aperfei\c{c}oamento de Pessoal de N\'{i}vel Superior - Brasil (CAPES) - Finance Code 001 and by the ELIOT project (ANR-18-CE40-0030 / FAPESP 2018/12579-7).
Gustavo A. Nunez Segura is supported by Universidad de Costa Rica. 
% \section*{References}
\bibliographystyle{IEEEtran}
\bibliography{bibliography}

\end{document}